# BioPhysical Modeling, Characterization and Optimization of Electro-Quasistatic Human Body Communication

Shovan Maity, Mingxuan He, Mayukh Nath, Debayan Das, Baibhab Chatterjee, *Student Member, IEEE* and
Shreyas Sen, *Senior Member, IEEE*
*School of Electrical and Computer Engineering, Purdue University*

*Abstract*—Human Body Communication (HBC) has emerged as an alternative to radio wave communication for connecting low power, miniaturized wearable and implantable devices in, on and around the human body. HBC uses the human body as the communication channel between on-body devices. Previous studies characterizing the human body channel has reported widely varying channel response much of which has been attributed to the variation in measurement setup. This calls for the development of a unifying bio-physical model of HBC supported by in-depth analysis and an understanding of the effect of excitation, termination modality on HBC measurements. This paper characterizes the human body channel up to 1MHz frequency to evaluate it as a medium for broadband communication. A lumped bio-physical model of HBC is developed, supported by experimental validations that provides insight into some of the key discrepancies found in previous studies. Voltage loss measurements are carried out both with an oscilloscope and a miniaturized wearable prototype to capture the effects of non-common ground. Results show that the channel loss is strongly dependent on the termination impedance at the receiver end, with up to 4dB variation in average loss for different termination in an oscilloscope and an additional 9 dB channel loss with wearable prototype compared to an oscilloscope measurement. The measured channel response with capacitive termination reduces low-frequency loss and allows flat-band transfer function down to 13 KHz, establishing the human body as a broadband communication channel. Analysis of the measured results and the simulation model shows that instruments with 50Ω termination impedance (Vector Network Analyzer, Spectrum Analyzer) provides pessimistic estimation of channel loss at low frequencies. Instead (1) high impedance (2) capacitive termination should be used at the receiver end for accurate voltage mode loss measurements of the HBC channel at low frequencies. The experimentally validated bio-physical model shows that capacitive voltage mode termination can improve the low frequency loss by up to 50dB, which helps broadband communication significantly.

## I. INTRODUCTION

Advancement of semiconductor technology over five decades following Moore's law has resulted in exponential reduction in the size and cost of unit computing. This has led to the development of small form factor wearable, implantable devices, which reside in and around the human body. These devices form a network of devices called the Body Area Network (BAN). The typical communication between BAN devices is through wireless communication using radio waves. However due to the proximity of these devices to the human body, the body itself can be used as the communication medium between these devices due to its good conductive electrical properties. This has led to the development of Human Body Communication (HBC). HBC based circuits and systems promises to be more energy efficient than wireless communication due to the relatively lower channel loss provided by the human body compared to the wireless media. Also, in HBC, the signal is primarily contained within the human body unlike radio wave communication, where the signal is radiated isotropically. Hence HBC provides enhanced security, as it is harder for an attacker to snoop the transmitted signal. These advantages make HBC a promising alternative to wireless radio communication for connecting small, energy constrained devices on and around the body and enable applications such as wearable health monitoring [1], information exchange between wearable devices [2], [3], secure authentication using a body worn key etc.

HBC can be divided into two categories: a) Capacitive, b) Galvanic, depending on the excitation, termination and how the return path is formed. In capacitive HBC [4], the signal is coupled through a single electrode into the body. The return path is formed through capacitive coupling of the receiver and transmitter ground with the earth ground and other surroundings. In Galvanic HBC [5], a differential signal is applied to the body through a signal and ground electrode. The resultant field is picked up at the receiver end as potential difference, through a pair of differential electrodes. In this paper we focus on the measurement of capacitive HBC characteristics.

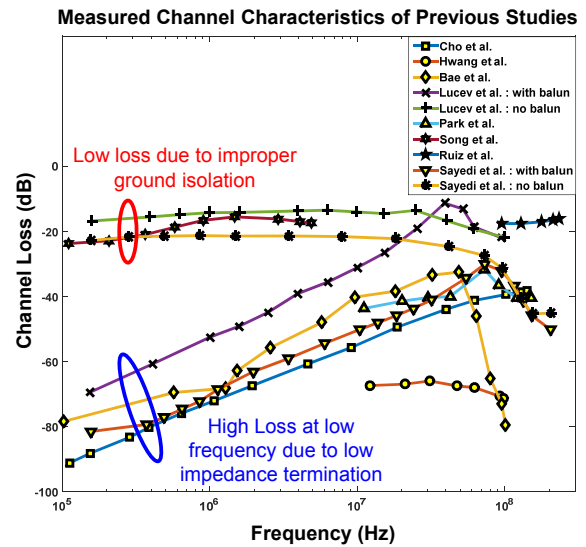

Figure 1: Channel Measurement plots from previous studies, showing wide variance in measured characteristics. Some of the measurements are carried out with transmitter and receiver having insufficient ground isolation. Some other studies achieve isolation through baluns but use Vector Network Analyzer/ Spectrum Analyzer for measurement, which has a low impedance termination, resulting in higher low frequency loss and hence a high pass response.

Proper understanding of the human body channel characteristics is necessary in designing efficient HBC transceivers. Previous studies [6]–[15], which characterizes the body channel has shown wide variation in the measured channel characteristics as shown in Figure 1. This can be attributed to two primary reasons: 1) insufficient ground isolation between the transmitter and the receiver, resulting in lower measured loss; 2) low impedance termination at the receiver end, resulting in higher loss at low frequencies. The devices communicating through HBC have small ground planes and is isolated from each other. So, in case of measurement devices with large ground planes, insufficient ground isolation between transmitter and receiver results in optimistic estimation of the channel loss. Similarly, source resistance at the transmitter end and termination impedance at the receiver end are key factors affecting the channel loss characteristics. This paper builds a lumped bio-physical capacitive HBC circuit model, which explains the experimental human body channel loss for frequencies up to 1MHz. The signal wavelength at these frequencies (>300 m) are orders of magnitude bigger than the human body size (2 m). The optimum signaling modality for voltage mode human body communication is also discussed with simulation/ experimental analysis done to understand the effect of each of those factors on the overall channel characteristics and explain some of the discrepancies seen in previous channel measurement studies. Following are the key contributions of this paper:

1. Developed a lumped bio-physical HBC circuit model, which explains the experimental channel loss characteristics up to 1MHz frequency for different excitation and termination. This also provides a unifying explanation to some of the measurement results shown in earlier studies.
2. Developed a miniaturized wearable prototype, which uses time domain sampling, for voltage mode channel measurements to replicate the scenario of an actual wearable to wearable communication.
3. Measured the human body channel loss at low frequencies (10 KHz – 1 MHz) to characterize it as a broadband communication channel, which can support up to 1 Mbps data rate.
4. Analyzed the effect of termination impedance (low vs high impedance, capacitive vs resistive) and source impedance (low vs high impedance) on the overall channel characteristic and provide recommendations for optimum HBC signaling.

The rest of the paper is organized as follows: Section II compares previous channel measurement studies in terms of the experimental setup used and the measured channel characteristics. The bio-physical model of HBC is developed in Section III. Section IV discusses the advantage of Broadband HBC and motivates the need for channel characterization in the low frequencies. Section V characterizes the human body forward path channel. Section VI discusses about non-common ground measurement setup with both an oscilloscope and a miniaturized wearable receiver. Section VII compares the non-common ground measurement results with the simulation results from the model. Section VIII looks into the optimum signaling modality for voltage mode HBC measurements and explains the reason behind some of the discrepancies seen in previous studies and Section IX concludes the paper.

## II. RELATED WORK

Several studies have been carried out to characterize the Capacitive HBC channel loss. However, these studies show a wide variance in the measured channel response. This shows a strong sensitivity of the measurement setup on the results. In this section we discuss about the experimental setup and measurement results of some of those studies.

Cho *et al*. [6] reported the human body characteristic in the 100KHz-100MHz range to be that of a band-pass filter with loss as high as 90dB at 100KHz. The pass band frequency is greater than 10MHz and the loss at the pass band varies between 35dB to 60dB depending on transmitter, receiver distance and also on the size of the ground electrodes. A battery powered signal generator with a programmable frequency synthesizer is used to transmit signal into the body. A grounded oscilloscope or spectrum analyzer is connected to the receiver electrode to measure the received signal. So in this case the transmitter and receiver are electrically isolated but the receiver has a strong ground connection.

Lucev *et al*. [7] uses a network analyzer to provide input to the transmitter electrode and also observe the received signal from the receiver electrode. But this provides a strong return path for the current from the transmitter to the receiver as their ground is common to that of the internal ground of the network analyzer. To circumvent this common ground issue, they introduce baluns at both the receiver and transmitter end to electrically isolate the two grounds. The channel characteristic without the baluns show a flat band response in the frequency range of 100KHz-100MHz, with a loss of around 20dB. Whereas introducing the baluns show a bandpass characteristic with loss of ~80dB around 100KHz and the minimum loss of ~20dB around 35MHz. The loss varies depending on the distance between the transmitter and receiver.

Bae *et al*. [8] also characterize the human body channel in the 100KHz-100MHz range with a battery operated transmitter and an oscilloscope isolated through a balun to observe the signals. The authors report a bandpass characteristic with a peak frequency of ~40-60MHz, which varies depending on the distance between the transmitter and receiver electrodes. The channel loss at a particular frequency is also dependent on the distance between the transmitter and receiver, but shows significantly higher variation than those reported in [7].

Hwang *et al*. [9] uses a battery powered transmitter and an oscilloscope at the receiver end, which are synchronized through an optical cable. They use a differential probe at the receiver end to isolate the receiver ground from the oscilloscope ground. Measurement results show that the human body channel shows an almost flat band response in the 10-100MHz frequency range, with loss of ~75dB with a peak of 5dB extra loss around 40MHz. The oscilloscope ground is then intentionally connected to the receiver ground and the measured characteristics show a 6dB change on average loss compared to the non-ground sharing case, but retaining the flat band response. This is contrary to [7] where the introduction of a balun to isolate the receiver ground changed the response from a flat band to a band pass response.

Ruiz *et al*. [10] carry out experiments in the 100MHz-1.5GHz frequency range with a network analyzer to provide and receive signals for different electrode configuration and sizes. The measured channel characteristics show quite a lot of variation depending on the electrode configuration. But all of them show a low pass response with a minimum loss of around 20dB at lower frequencies for a particular configuration.

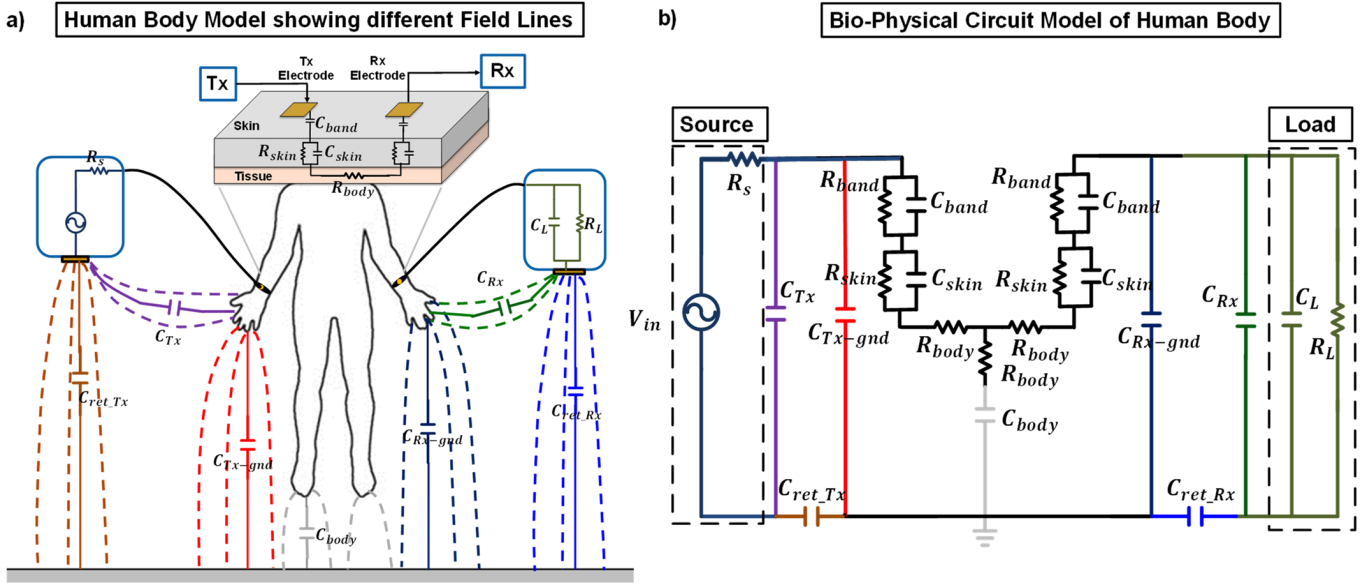

Figure 2: a) Diagram explaining the origin of different resistances and capacitances of the HBC model. The parasitic capacitances are formed by coupling between the body, transmitter/receiver ground plane and the earth ground. The internal body resistance and capacitances are shown in the tissue model. B) The HBC circuit model considering all the components as in Figure 2a.

Callejon et al. [11] carry out experiments for both galvanic and capacitive HBC and study the effects of channel length, electrode type, different human subjects and different parts of the body, posture of the subject on the channel characteristics. The capacitive coupling measurements are done at 1-100MHz, with input and output both provided by spectrum analyzer and isolated by baluns at both sides. The channel response is bandpass, with peak between 60-70MHz. The channel loss has weak dependence on channel length, with a difference of around 10dB for channel lengths of 15cm and 125cm. Configurations with large ground electrode sizes showed lesser loss. Postures, where the subject was sitting, showed lesser loss than postures when the subject was standing/walking. The authors attribute it to lesser distance between external ground and ground electrodes in sitting posture for the lesser loss.

Callejon *et al*. [12] carry out experiments with different kind of transmitter and receiver ground connections in 10KHz-1MHz frequency range to study the effect of experimental setup on measured characteristics of the human body channel. They carry out experiments in 5 different setups: receiver, transmitter ground connected; transmitter ground isolated through a balun keeping the receiver connected to ground; transmitter connected to ground keeping receiver isolated through balun; both receiver and transmitter isolated through balun; battery operated receiver and transmitter with no connection to earth ground. The experimental results show that the results from the battery operated scenario matches closely with the case where only the transmitter is isolated through a balun and the loss characteristic is almost flat over the frequency range with loss of ~20dB. This matches closely with the simulation model that has been built in the paper. Whereas their experiments showed a band pass response, while introducing a balun in the receiver side irrespective of the transmitter being connected to ground or not and this deviates from the simulation results of the model they have developed. From their experiments the authors conclude to discard any measurement setup that introduces baluns at the receiver side. But this kind of setup has been used in [7], [8].

Park *et al*. [13] characterize the human body channel through a setup containing miniaturized wearable devices. The channel is characterized in the 20-150MHz frequency range. The response shows a peak around 70 MHz with a loss of around -35dB. The overall channel response in this range is almost flat with loss ranging from -42dB to -32dB. The authors suggest using a miniaturized wearable setup for measuring channel characteristics and show that measurement setup using VNA or spectrum analyzer provide optimistic channel loss measurements even with baluns used for isolation.

The studies conducted so far either characterize the human body as a bandpass or a lowpass channel with significant variation in the loss magnitude. The goal of this paper is to explain the sources of discrepancy/ variation in the measurements in literature and provide a unifying explanation and repeatable channel models. We carry out measurements up to the frequency range of 1MHz with voltage mode signaling, which will enable characterizing the human body as a channel for broadband communication supporting data rates up to 1Mbps.

III. BIO-PHYSICAL HBC CIRCUIT MODEL

*A. Human Body Forward Path Components*

In HBC the transmitter couples the signal through a metal electrode onto the skin. The outer most layer of the skin has higher impedance compared to the inner layer of fat, tissue and conductive fluids [14]. So, the signal transmission is primarily through the internal layers of the body. At the receiver end similar electrodes are used to receive the signal. As shown in Figure 2a, the electrode to skin contact can be modeled with a capacitance ($C_{band}$). A tighter contact will result in higher capacitance and hence lesser impedance. The capacitance value is dependent on the separation and is in the order of 100s of pF if the electrode skin contact distance <0.1mm. The skin impedance is also shown to be in the order of KΩ as found by previous studies [5], [18]–[20]. The internal fat and conductive tissue has impedance in the order of 100s of ohms [18], [21]. The source impedance of the transmitter is a few ohms. The load impedance provided is determined by the receiver device. If an oscilloscope is used for measurements, the impedance provided by the probe is a combination of resistance (1MΩ, 10MΩ) and capacitance (13pF, 79pF). These values are used to construct the circuit model of HBC transmission as shown in Figure 2b. The circuit model is also dependent on the termination and excitation modalities. For example, a single ended excitation model will have one electrode and skin impedance connected to the source, whereas a differential excitation will have two electrode impedances and two skin impedances connected to the source. The values of each of the resistance and capacitances used in the HBC bio-physical model and the parameters considered to get those values are discussed in Table I.

TABLE I: HBC Bio-Physical Circuit Model components

| Component | Value | Comments |
|---|---|---|
| **Source Resistance ($R_s$)** | 50Ω | Source impedance of the signal generator |
| **Band to Skin Capacitance ($C_{band}$)** | 200pF | The transmitter/receiver electrode is in direct contact with the skin for our experiments, hence the air gap (d) between them is very small (<0.1mm). With an electrode size of 4cm² and d=0.01mm, C=354pF. This capacitance will become prominent when the electrode skin contact is loose. |
| **Band to Skin Resistance ($R_{band}$)** | 100Ω | Tight contact between skin and electrode will create a resistive path between them with low resistance. |
| **Skin Layer Resistance ($R_{skin}$)** | 10KΩ | The skin impedance varies in the range of 1KΩ-100KΩ depending on skin moisture and other factors [18]–[20]. |
| **Skin Layer Capacitance ($C_{skin}$)** | 90pF | Typical skin layer thickness is 0.2-4mm, hence taking a maximum thickness (d) of 4mm, dielectric constant of ~100 [20], skin area (A) of 4 cm² near the transmitter or receiver, the calculated capacitance is 90pF ($C=100*\epsilon_o*A/d$) |
| **Tissue Resistance ($R_{body}$)** | 200Ω | Subcutaneous tissue resistance is 100Ω-400Ω [18], [21] |
| **Feet to ground capacitance ($C_{body}$)** | 9pF | Assuming a feet area of ~100 cm² and feet to ground distance of 1cm, calculated capacitance is around 9pF |
| **Body to ground capacitance at Transmitter ($C_{Tx-gnd}$)** | 75pF | Half of the capacitance value determined from the experiment described in Section III.C |
| **Body to ground capacitance at Receiver ($C_{Rx-gnd}$)** | 75pF | Half of the capacitance value determined from the experiment described in Section III.C |
| **Transmitter ground to body capacitance ($C_{Tx}$)** | 300fF | Depends on the ground plane size and separation of ground plane from body. Ground plane size of 4 cm² and separation of 1cm body results in a cap of 354fF |
| **Receiver ground to body capacitance ($C_{Rx}$)** | 300fF | Calculated in the same way as transmitter to ground capacitance, added as part of load capacitance as value is small compared to the load capacitance |
| **Transmitter Ground to earth Capacitance ($C_{ret\_Tx}$)** | 1.5pF | Experimentally determined by experiment in Section III.D |
| **Receiver Ground to earth Capacitance ($C_{ret\_Rx}$)** | 1.5pF | Experimentally determined by experiment in Section III.D |
| **Load Capacitance ($C_L$)** | 13pF for 10x probe 79pF for 1x probe 1pF for wearable | The load capacitance for the probes are taken from the data sheet. The input capacitance of the wearable device will be lower due to its smaller size. |
| **Load Resistance ($R_L$)** | 10MΩ for 10x probe 1MΩ for 1x probe 10MΩ for wearable | Load resistance values for the probes are taken from the data sheet. Since the wearable IC is made in CMOS technology, its input impedance is primarily capacitive and hence the input resistance is taken as 10MΩ (high value). |

*B. Return Path Capacitance Estimation experiment*

The overall channel loss is strongly dependent on the return path capacitance between the transmitter and the receiver. To estimate the return path capacitance, several known value capacitances ($C_{expt}$) are connected between the transmitter and receiver ground and the loss value measured. The experimental setup and the equivalent circuit model is shown in Figure 3. From the circuit model in Figure 3b, the channel loss can be estimated as $\frac{C_g}{C_L+C_g}$, which is the capacitive division ratio of the equivalent return path capacitance ($C_g$) and the load capacitance ($C_L$). Now the return path capacitance in presence of the extra experimental capacitance is $C_g = C_{expt} + \frac{C_{ret}}{2}$, assuming $C_{ret\_Tx} = C_{ret\_Rx} = C_{ret}$. From the loss measurements with different $C_{expt}$, we can now estimate the return path capacitance

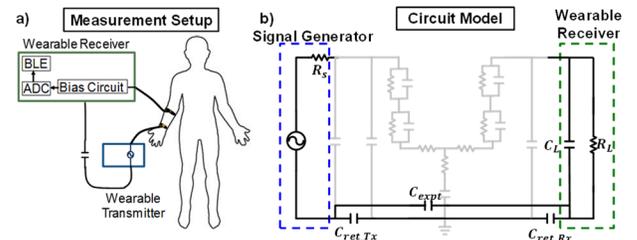

Figure 3: a) Measurement setup used to estimate the return path capacitance between a wearable transmitter and receiver. b) Circuit model of the measurement. The equivalent return cap is $C_g = C_{expt} + \frac{C_{ret\_Tx}*C_{ret\_Rx}}{C_{ret\_Tx}+C_{ret\_Rx}}$. Loss measurements with multiple known value capacitances ($C_{expt}$) enable us to find the unknown return path capacitances ($C_{ret\_Tx}$, $C_{ret\_Rx}$).

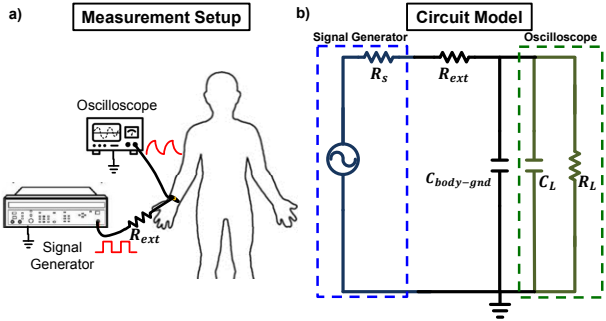
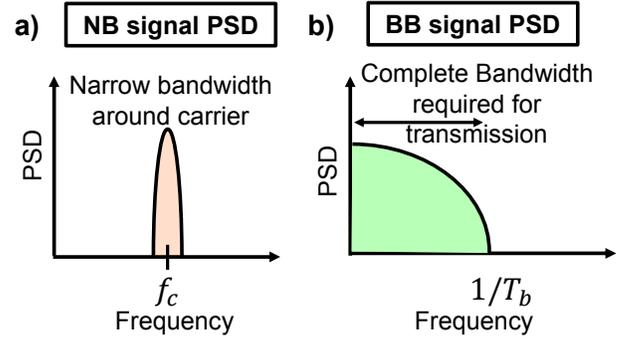

Figure 4: a) Measurement setup used to estimate the capacitance between the body surface and ground. The input signal is applied to the body through a series resistance and observed through a oscilloscope. b) The equivalent circuit diagram of this experiment. The body to ground capacitance will act as an extra parallel capacitance to the oscilloscope load capacitance and hence affect the time constant of the received signal.

Figure 5: a) Power Spectral Density (PSD) of a broadband transmitted signal occupying the complete bandwidth from DC up to data rate, motivating the characterization of the human body channel up to low frequencies. b) PSD of a narrowband signal occupying only a small bandwidth around the carrier frequency, requiring the channel to be characterized around the carrier frequency and not until low frequencies.

($C_{ret}$). The estimated return path capacitance from this experiment is around 1.5pF. We use this value of return path capacitance for the simulations of the HBC circuit model as shown in Table I.

*C. Skin to Ground capacitance estimation experiment*

One of the key components of the proposed HBC circuit model is the capacitance between the human body and the earth ground. The experimental setup shown in Figure 4a is used for this measurement purpose. A signal generator is used to provide square wave input and is applied to the body through a series resistance. The series resistance is used to increase the time constant of the received signal at the receiver end. First the time constant of the signal is measured by directly applying it to the oscilloscope. The time constant in this scenario ($\tau_{osc}$) will be dependent on the series resistance and oscilloscope load capacitance, $\tau_{osc} = R_{ext} C_L$. Now to measure the effect of capacitance between the body and external ground, the signal is applied to the body and the time constant measurement repeated. The time constant in this scenario ($\tau_{osc-body}$) is dependent on both the oscilloscope load capacitance as well as the body to ground capacitance, $\tau_{osc-body} = R_{ext}(C_{body-gnd} + C_L)$ (Figure 4b). Measurement results with two different series resistance shows the body to ground capacitance to be around 150pF. Since this consists of the total capacitance from the body to ground, during modeling we model half of the capacitance at the transmitter end and the other half at the receiver end.

### IV. BROADBAND HBC

In this paper, the HBC channel characterization is carried out in the frequency range of < 1MHz. The goal is to analyze the feasibility of using the body as a wire-like broadband communication channel. A broadband channel with 1MHz bandwidth can enable transmission up to 1Mbps data rate, which is sufficient for most HBC applications such as physiological health monitoring, remote authentication, exchange of business card / social networking request [1]–[3]. During transmission, broadband HBC does not require modulation and demodulation, which requires high power circuits. Also, broadband HBC utilizes the complete bandwidth of the body for data transmission. Hence Broadband HBC promises to be more efficient than narrowband HBC in terms of required energy per bit with achievable energy efficiency of up to 6.3pJ/bit [22]. In a narrowband scenario (Figure 5a) power can also be transferred for communication and hence the channel transfer characteristics can be determined as the ratio of received and transmitted average power [13]. However, broadband communication (Figure 5b) requires the signal to be sent as a voltage corresponding to a 1/0 bit. Hence to characterize the channel for possible usage in broadband HBC we use voltage mode signaling where voltage signal is applied as input. The received signal is also in voltage domain and is detected through time domain sampling of voltage signal. Due to the potential advantages of Broadband HBC, we characterize the HBC channel until low frequencies (10s of KHz) and compare the results with the simulation results from the developed bio-physical model in the following sections.

### V. COMMON GROUND MEASUREMENTS: HBC FORWARD PATH CHARACTERIZATION

Recently there has been studies [23] on characterizing the human body forward path channel loss. In this section we discuss about the forward path characterization experiment and compare the measurement results with the results obtained from the bio-physical model to validate the model.

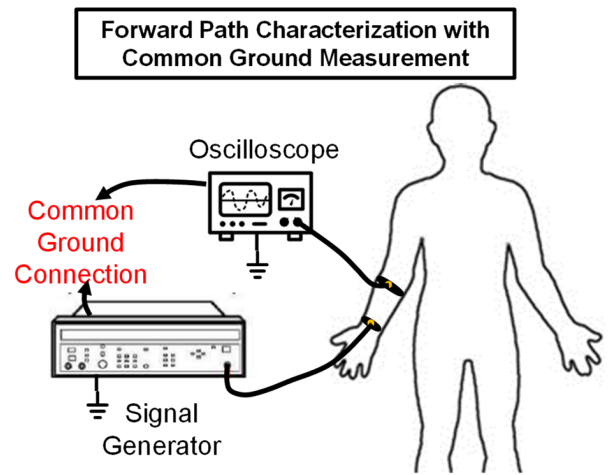

Figure 6: Measurement setup for human body forward path characterization. A signal generator is used as a transmitter and an oscillscope measures the received voltage. The transmitter and receiver grounds are connected to remove the effect of return path capacitance on the measurement and characterize only the forward path

## A. Motivation

The channel loss in HBC has two primary sources: 1) The transmission channel, 2) The return path. To design better interface circuits with the human body channel, in order to minimize total loss, it is necessary to characterize and decouple the major sources of loss, e.g. if most of the loss is coming from the body or from the capacitive return path. In this section we try to address this issue by doing common ground measurements and hence removing the non-common ground effect. This will enable us to characterize the human body channel forward path and also determine the values of the different circuit elements used to characterize the human body.

## B. Experiment Setup

A BK precision 4055 signal generator is used to provide a sine wave input at the TX and the signal at the RX end is observed in a Tektronix DPO 7104 oscilloscope (Figure 6). The oscilloscope and the signal generator grounds, are shorted to eliminate any effect coming from the return path through the power supply. Copper electrodes are used to couple the signal into the body. The frequency of the input sine wave is changed and the attenuation at the receiver end is measured to find the human body channel loss characteristics over different frequencies. Experiments are carried out for both differential (DE) and single ended (SE) excitation at multiple frequencies to find the forward path channel characteristics of the human body. The circuit model for each of these scenarios is shown in Figure 7. Voltage mode signaling is used for the experiments, which is provided by a low output impedance transmitter and a high input impedance receiver.

## C. Measurement Results

Experimental results (Figure 8) show that single ended excitation and termination shows minimum loss ~0dB. Differential excitation with single ended termination, on the other hand shows a loss of ~6dB whereas differential excitation and termination results in ~10dB loss. This shows that the forward path channel loss not only depends on the human body channel characteristics but also on the excitation termination modalities. The measured loss characteristics match with the simulation model with the values of circuit components as shown in Table I.

## D. Conclusion: Channel Loss dominated by Return Path

From the experimental results, with common ground it is clear that forward path channel loss of the human body is small and most of the loss in HBC is due to the non-common ground return path capacitance. Also, excitation termination modality has direct effect on the forward path loss, hence the receiver and transmitter configuration and impedance values have to be chosen properly to minimize the channel loss.

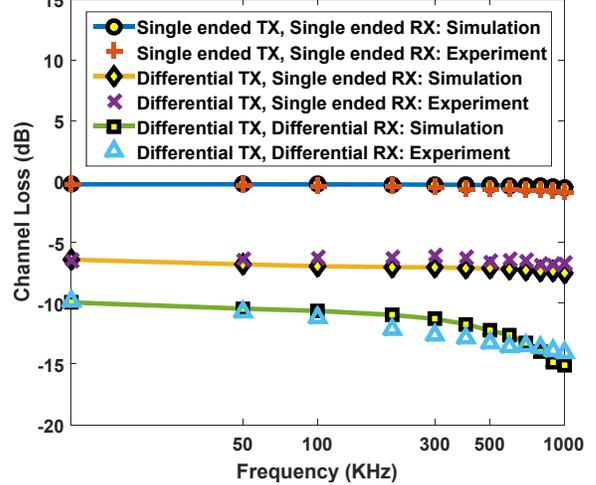

Figure 8: Measurement results showing the characteristics of human body forward path under different excitation, termination modalities. The measured channel loss show close correspondence with the simulated loss from the developed human body circuit model.

## VI. NON COMMON GROUND: MEASUREMENT SETUP WITH CAPACITIVE RETURN PATH

This section discusses the measurement set up used for measuring the human body channel loss in a non-common ground scenario and experimentally validate the different component values of the human body model. We use two different type of measurement setup for the channel measurements: (1) An oscilloscope based measurement to provide high impedance capacitive termination for voltage mode channel measurements, (2) a miniaturized wearable receiver to replicate an actual HBC scenario without any measuring instruments with large ground. Recently [13] has shown a wearable measurement setup for power loss measurement. Here we carry out wearable measurements to 1) perform voltage loss measurements in the low frequency range to characterize it for time domain broadband communication, 2) validate the developed bio-physical model for wearable scenarios. A battery operated signal generator is used to provide voltage signal as stimulus. The signal is coupled to the body through copper electrodes connected to a band. The measurement setup and the circuit model of the system with oscilloscope and wearable receiver is discussed in the next subsections.

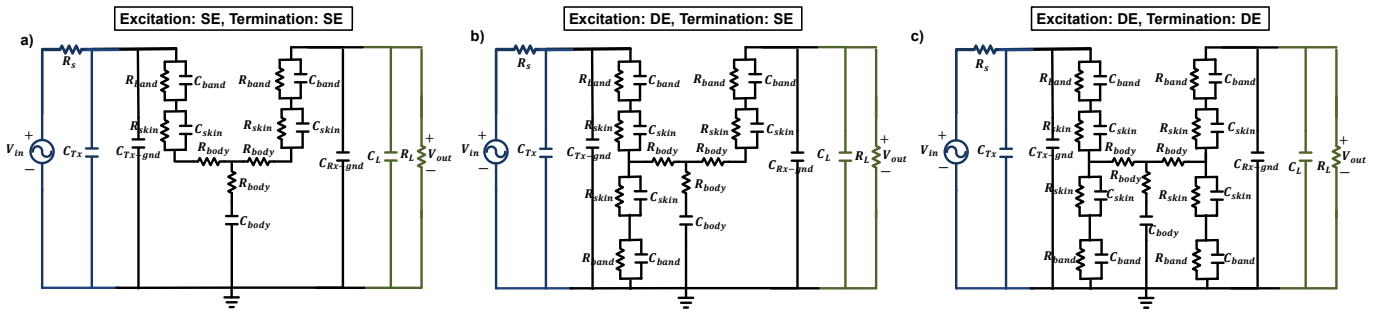

Figure 7: Lumped Circuit Model of the human body for different transmitter receiver electrode configuration for forward path characterization with common ground between transmitter and receiver. Circuit model with: a) Single Ended (SE) excitation and Single Ended (SE) termination, b) Differential Ended (DE) excitation and Single Ended (SE) termination, c) Differential Ended (DE) excitation and Differential Termination. These different excitation/termination modalities help validate the developed circuit model and the component values.

### A. Oscilloscope based Setup: Large Ground

To achieve high impedance termination for voltage measurements an oscilloscope is used as the receiver (Figure 9a). The transmitter consists of a battery operated miniaturized signal generator. The termination impedance provided by the oscilloscope is dependent on the probe used for measurement. In our experiments we use two different probes to provide capacitive and resistive termination of different values. This helps validating the developed bio-physical HBC circuit model parameters under these scenarios. Since we want a high impedance termination at the receiver end an oscilloscope is used for voltage measurement. Figure 10a shows the circuit model with a 10x probe with an equivalent resistance of 10MΩ and capacitance of 13pF. The 1x probe provides an equivalent resistance of 1MΩ and capacitance of 79pF as shown in Figure 10b. Both the probes provide a high impedance capacitive termination at the receiver end. However, since the oscilloscope is a wall power supply connected device with a large ground plane, the return path capacitance between the transmitter and the receiver will be larger than that of a wearable scenario and result in an optimistic measure of the HBC channel loss. Hence it is required to build a miniaturized receiver device, which will provide a more accurate estimation of the HBC channel loss.

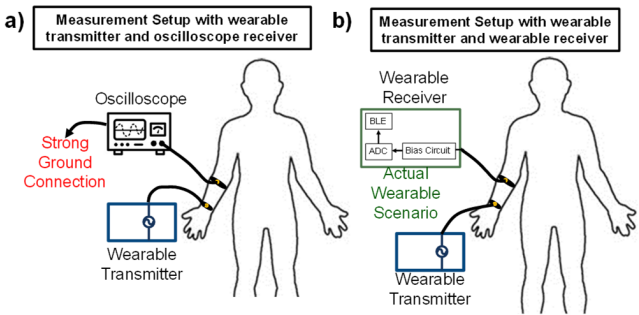

### B. Miniaturized Wearable Protoype based Setup

The oscilloscope measurements discussed in the previous subsection do not emulate an actual wearable scenario due to the presence of a large ground at the receiver end. We use a battery-operated device, which has a small ground plane, at the receiver end to emulate a more realistic HBC scenario (Figure 9b). Since our experiments are carried out in voltage mode, the received signal is a voltage waveform and needs to be sampled in time domain for measurement. Hence, we use an ADC at the receiver end to sample, digitize the received waveform and recover the received signal amplitude information. For the miniaturized receiver we use a STM32F103C8 microcontroller, which has 12 bit, 1 MSPS ADC modules. The sampled ADC data is stored in the microcontroller at the receiver end and then transmitted through an HC-06 Bluetooth module. The transmitted data is acquired through a similar HC-06 Bluetooth module connected to the PC and processed in MATLAB to extract the received amplitude information. The conceptual block diagram of the receiver and the data acquisition system is shown in Figure 11a and Figure 11b respectively. Figure 11c shows the actual implementation of the miniaturized receiver device.

The maximum frequency signal which can be digitized to recover amplitude information through sampling is limited by the sampling rate of the ADC. With the ADC sampling rate limited to 1MSPS, a 10x oversampling will limit the maximum frequency of operation to 100 KHz. So to characterize the channel up to 1MHz with a 1 MSPS ADC a histogram based approach is used. In this approach a square wave is used as the input waveform. The signal is sampled randomly and the difference between two consecutive samples is measured. The whole amplitude range is divided into a few smaller bins and the histogram of the difference is computed. For an ideal square wave input the difference will either be zero or equal to the amplitude of the signal. Hence if the histogram of the difference between two consecutive samples is plotted there will be two sharp peaks around 0 and the amplitude of the signal as seen in Figure 12. Hence by analyzing the histogram of the sampled signal it is possible to estimate its amplitude. Since the sampled voltage can only be on either of the two levels of the square wave, it is not required to reconstruct the signal accurately in time domain to estimate its amplitude. Hence, this approach is not dependent on the sampling rate and will provide accurate results even if the sampling rate is lower than the frequency of the received signal. So even by subsampling the signal it is still possible to find the amplitude information, enabling measurements up to 1MHz with a 1MSPS ADC. In a real scenario, due to presence of noise, the received signal will not be a perfect square wave and the histogram peaks will be spread out. So for our experiments the histogram is taken multiple times and averaged to find the average peak and subsequently the amplitude of the signal.

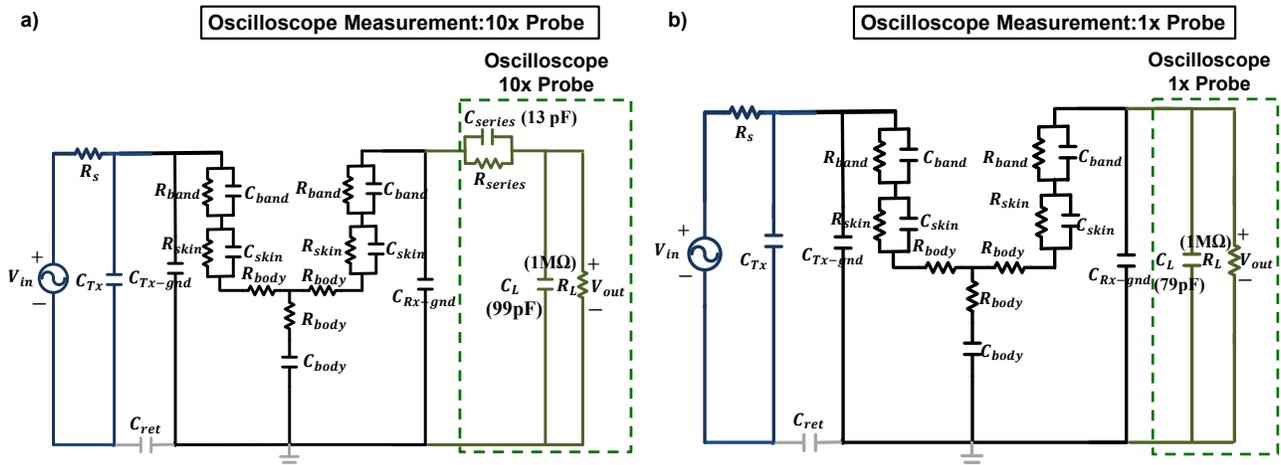

Figure 10: Human body circuit model with an oscilloscope as a load device at the receiver end. a) Oscilloscope with a 10x probe provides an equivalent load of 10MΩ resistance in parallel with a 13pF capacitance. b) An 1x probe at the oscilloscope provides a load with 1MΩ resistance in parallel with 79pF load. The HBC model is used to provide loss estimates for both these load scenarios and compared with the experimental measurements.

Figure 9: a) Measurement setup using a wearable signal generator as transmitter and oscilloscope as receiver. The oscilloscope is a ground connected instrument and hence do not represent an actual scenario for werable HBC measurement. b) Setup using a wearable transmitter and wearable receiver, consisting of a bias circuit, ADC and bluetooth module. This represents an actual wearable HBC scenario

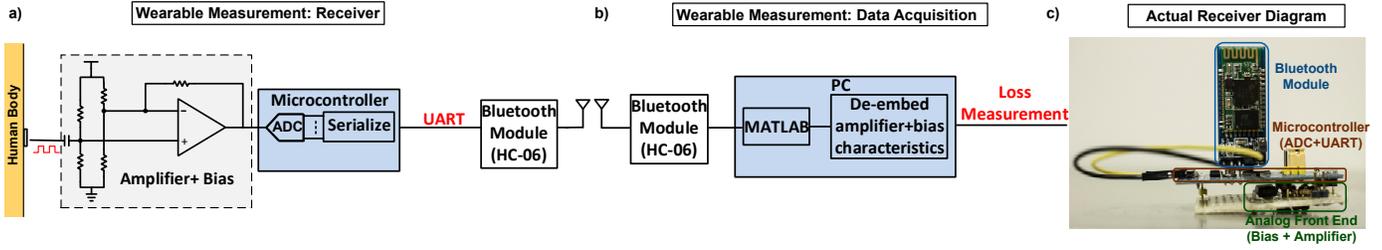

Figure 11: a) Block diagram of the receiver used for wearable measurements. The received signal is amplified with a known gain amplifier and then digitized using an ADC, which sends the bits out through Bluetooth after serializing them. b) Setup to acquire the data sent from the receiver and do the loss measurements. The data transmitted from the wearable receiver is processed in MATLAB after de-embedding the circuit characteristics and plotted to find the channel characteristics with miniaturized wearable transmitter and receiver. c) Actual diagram of the wearable receiver device showing the different components.

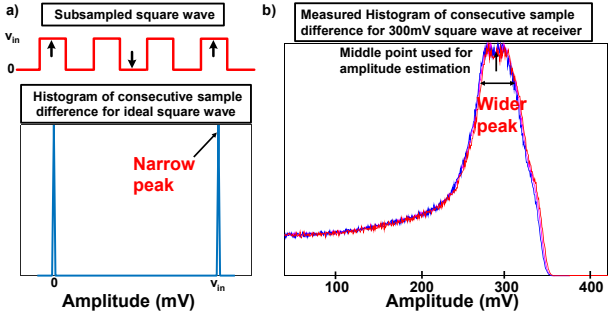

Figure 12: a) Histogram plot of difference between consecutive samples from a sub sampled ideal square wave. b) Actual measured histogram plot for a 300mV received signal through HBC. The plot only shows the peak corresponding to amplitude of the signal. Since the peak shows a wide spread, the mid-point is taken as the amplitude.

Since the wearable transmitter and receiver do not have a common ground reference, it is necessary to bias the signal at the receiver end before applying it to the ADC. This requires a resistive bias circuit as shown in Figure 11a. Also we amplify the biased signal through a known gain amplifier and apply it to the ADC. This amplifies the received signal but do not amplify the noise in the ADC, which enables detecting the received signal correctly. Hence for channel characterization we need to de-embed the amplifier characteristics from the received signal. In this scenario we have an amplifier of gain 12 over the measured frequency range, hence the received signal amplitude is divided by 12 to find the channel loss. Also the bias circuit for the amplifier creates a high pass response which needs to be de-embedded from the measured response to get the actual channel response. Hence the final response is obtained by de-embedding the amplifier and bias circuit response from the measured channel characteristics. The measurement results from these two different experimental setup is discussed in the next section. The equivalent circuit diagram of this setup is shown in Figure 13.

### VII. Measurements: Broadband HBC

This section discusses the measurement results through oscilloscope and wearable receiver and compares them with the developed circuit model. The experiments were carried out on 4 male subjects over 6 days. The transmitter was placed on the forearm and the receiver is placed on the wrist. The resulting HBC channel length is 20-25 cm.

#### A. Oscilloscope based Measurement Results

Two different probes are used for oscilloscope measurements providing termination with different capacitances. It is expected that a capacitive termination will result in a channel loss determined by the ratio of return path capacitance and load capacitance and hence the response should be constant over frequency. The measured channel response is indeed flatband, as seen in Figure 14. This matches closely with the simulated results from the model. Also, it is expected that a higher termination capacitance should result in a higher loss. As seen from Figure 14, the channel loss during measurement through a 1x probe (79pF capacitance) is around -47 dB. This is higher than the -43 dB channel loss measured with 10x probe (13pF capacitance). These measurements show that it is indeed possible to have a flat-band channel until low frequencies by utilizing a capacitive termination at the receiver end. This shows the possible utilization of the human body channel as a broadband communication medium.

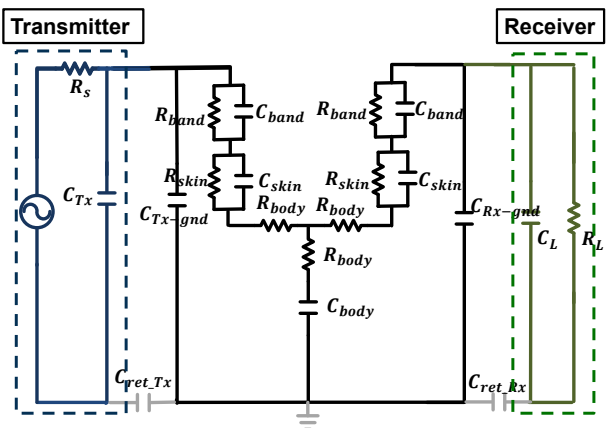

Figure 13: Circuit model for measurement setup with a wearable transmitter and receiver. The wearable load is represented as a parallel combination of a resistance and capacitance corresponding to the input impedance of the receiver.

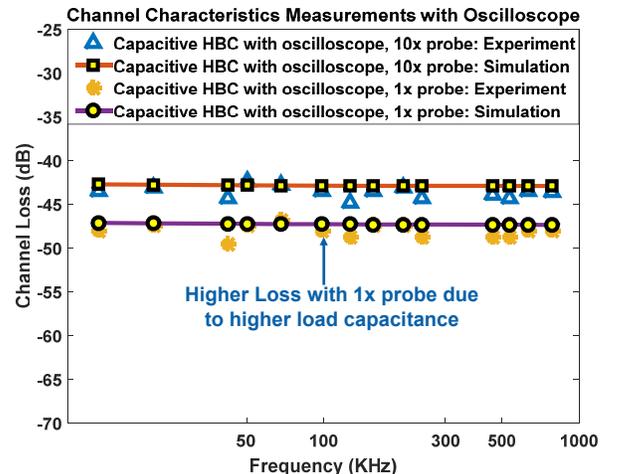

Figure 14: Comparison of oscilloscope measurement results for two different probes with simulation model. Simulation results correspond to the measurement results closely, showing higher loss for a 1x probe due to its higher load capacitance.

### B. Miniaturized Wearable Protoype based Measurement Results

The wearable measurements require an amplifier and a bias circuit at the receiver end. Since these circuits have their own frequency response, it is necessary to de-embed them from the measured characteristics. The amplifier has a flatband gain of 12 over the frequency range. The bias circuit has a high pass frequency response, which results in an overall high pass channel response. Hence de-embedding the bias circuit response from the overall response results in a flat band channel response as shown in Figure 15. However, the channel loss in a wearable scenario is higher (-49 to -52 dB) than oscilloscope measurements (-43 dB and -47 dB respectively with 10x and 1x probes). This is due the lower return path capacitance in this wearable measurement scenario, as there is no device with big ground. This shows that channel measurements with oscilloscope provides optimistic estimation of the channel loss and hence should be avoided.

### VIII. Optimum HBC Signaling Modality: Key Learnings

From the experimental results in Section III and Section VII it can be seen that the human body channel loss can be minimized by applying proper value and configuration of excitation and termination. In this section we discuss about some of the excitation and termination modalities that should be used for HBC channel measurements in voltage mode.

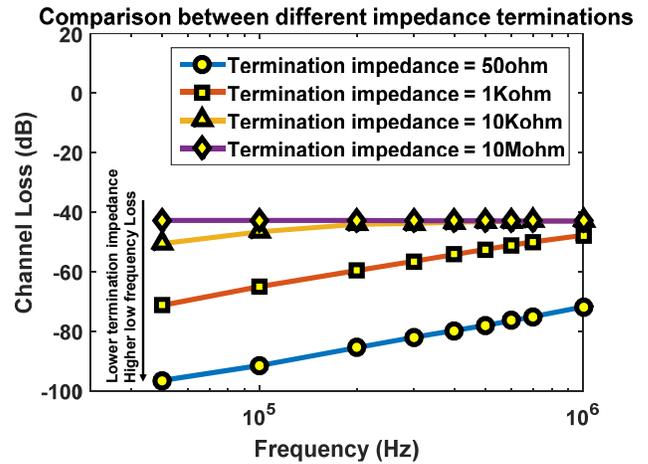

Figure 16: Comparison of channel characteristic for different termination impedance from simulation model. Lower termination impedance results in higher low frequency loss, as well as higher cut-off frequency. Hence it is necessary to have a high termination impedance receiver to measure channel characteristics at low frequencies.

### A. Avoid 50Ω Termination

Most previous studies for HBC channel measurements use a Vector Network Analyzer (VNA) or a Spectrum Analyzer to measure the received signal. Both of these instruments have a 50Ω input impedance and hence they provide a very low impedance termination at the receiver end. Since the human body channel impedance is of the order of a few KΩs, any voltage loss measurement using a 50Ω termination resistance will result in a high loss. This can be observed from Figure 17 where previous studies using VNA show about 20dB loss even when there is no ground isolation. Using a 50Ω termination in the developed HBC circuit model also shows similar loss. So to remove the effect of measurement setup on the channel loss a high impedance termination should be used at the receiver end. Hence a Vector Network Analyzer or a Spectrum Analyzer will provide a pessimistic estimate of the channel loss and should not be used in voltage mode channel loss measurements.

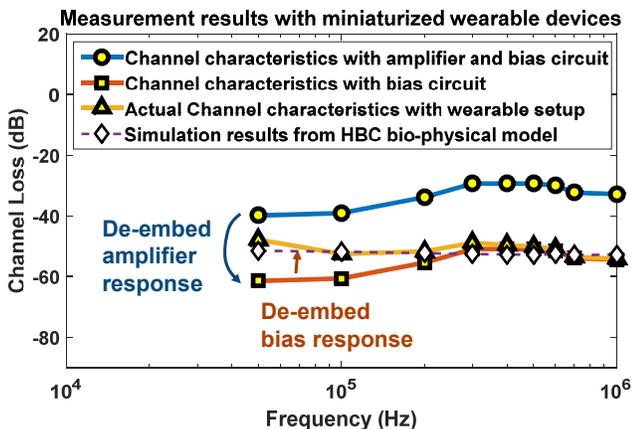

Figure 15: Measurement results using miniaturized wearable prototype. After de-embedding the amplifier characteristics and the bias circuit characteristics from the measured result the channel response is flatband. This matches closely with the simulation results assuming a 1pF input capacitance of the wearable receiver.

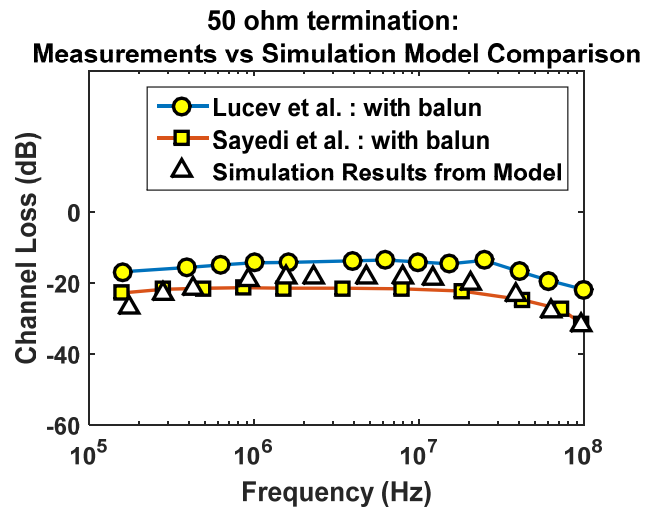

Figure 17: Comparison of simulation and measurement result from previous studies for a 50Ω termination at the load end. The measurements are carried out by simultaneously providing input and measuring the received signal in a VNA without using any balun for isolation, effectively making it a common ground measurement. Simulation results with common ground and 50Ω termination also show similar loss.

## B. Desired: High Impedance Termination

As discussed in previous literature and also found from the models in Section III, the human body impedance is in the order of a few KΩs. So, any voltage measurement with a low input impedance device will result in a higher loss, since most of the voltage drop will happen across the body. Also, the low termination resistance along with the return path capacitance will create a pole in the channel transfer function, resulting in a high pass response. Lower the termination resistance, higher the pole frequency, more the effect on low frequency measurements (Figure 16). Majority of the measurements in previous literature [6], [7], [11], [15] has been carried out with network analyzers or spectrum analyzers, which has 50Ω input impedance. This results in pessimistic measurement of channel loss. A high input impedance measurement device, such as an oscilloscope will provide a more realistic loss scenario, as it will replicate a CMOS receiver circuit with very high input impedance. So voltage loss measurements should be carried out with high impedance devices, which shows lesser loss than 50Ω termination measurements.

## C. Capacitive Termination: Low frequency loss reduction

The return path in capacitive HBC is formed by the coupling capacitance between the transmitter and receiver. Any resistive termination will result in high loss at lower frequencies, because the high impedance of the capacitor will result in most of the voltage drop happening across it. This results in a high pass response with cut-off frequency determined by the return path capacitance ($C_{ret}$) and termination resistance ($R_{load}$). The cut-off frequency ($1/R_{load}C_{ret}$) is inversely proportional to the load resistance and hence very low impedance termination (50Ω), as is the case for a spectrum analyzer or VNA, will result in a high cutoff frequency. So the low frequency measurements with 50Ω input impedance measurement devices will result in higher loss. On the other hand, a high impedance receiver `with capacitive termination will result in a flat-band channel response as the output voltage will be a capacitive division between the return path and termination capacitance. It is possible to achieve up to 50dB loss reduction at low frequencies through capacitive termination, as seen from Figure 18. This is also a realistic scenario for a CMOS device, where the receiver input capacitance will act as the termination capacitance and hence will enable a flat band channel until low frequencies. This explains the high pass response reported in many of the previous studies [6]–[8], [15] and also shows the feasibility of broadband transmission, using the human body as a communication medium.

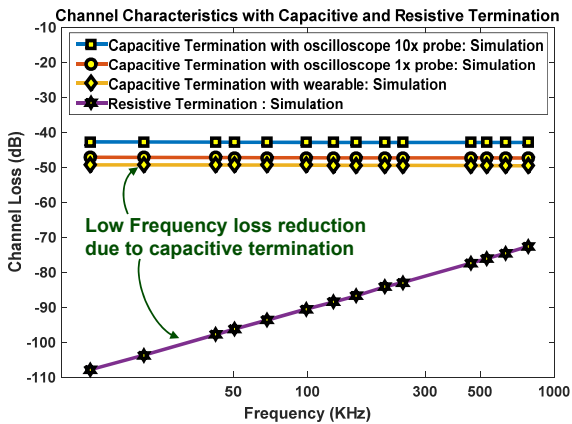

Figure 18: Simulation results showing the effect of capacitive termination on low frequency loss compared to a resistive termination. Capacitive termination results in a flat-band response of the human body channel. This also explains that, previous studies using VNA with 50Ω termination impedance for measurement at the receiver end, report a high pass response of the human body channel.

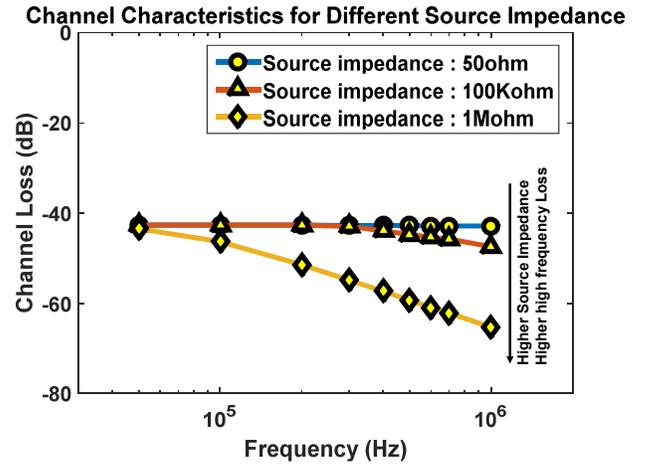

Figure 19: Comparison of simulated channel characteristic for different source impedance. Higher source impedance results in more high frequency loss and lower roll off frequency. So a low source impedance is desired at the transmitter end for voltage mode channel measurements.

## D. Voltage-Mode Signaling

Most previous studies use voltage signal as excitation and the channel loss is measured in terms of the ratio of received and transmitted voltage. Since voltage is the measured metric in this scenario, the signaling modality should maximize the received voltage. To that end, the receiver impedance can either be conjugate matched or be as high as possible. Conjugate matching is important for reducing reflections at the receiver end. However, at low frequencies (order of few MHz) the wavelength of the transmitted waves are significantly larger than the communication channel length. Hence it is not necessary to provide conjugate matching at this frequency range. So, the termination impedance has to be large for voltage mode signaling. The source impedance on the other hand has to be minimized to reduce voltage drop across it and maximize the voltage received at the receiver end. As can be seen from Figure 19, as the source impedance increases, the voltage loss increases for higher frequencies. Hence, it is desirable to have a low source impedance for voltage mode signaling.

## IX. CONCLUSION

This paper characterizes the human body channel up to 1MHz frequency for capacitive HBC, where the closed loop path is formed by return path parasitic capacitances. A *unifying* bio-physical model of HBC is developed with component values determined through experiments and previous literature. Common ground measurements with different excitation and termination modalities show that the loss through the human body forward path contributes very little to the overall channel loss and validates the forward path components of the HBC model. Non-common ground experiments with oscilloscope show almost constant loss down to low frequencies (10 KHz), highlighting the feasibility of using the human body as a broadband channel. The channel loss shows dependence on the termination impedance varying from -43dB to -47dB for 13pF and 79pF termination respectively. Wearable prototype based measurements in voltage mode with a sampling receiver has almost constant -49dB to -52dB loss across this frequency range, showing the need for actual wearable based measurements for accurate channel loss estimation. Analysis of the developed bio-physical model shows that *high termination impedance* and *low source impedance* provides the optimum channel loss and explains some of the high-loss measurements at these low frequencies in previous studies. These results show that the human body can be used as a broadband communication medium, *like a 'lossy wire'*, which can enhance the battery life of small form-factor energy constrained wearable/implantable devices opening up applications like remote health monitoring, secure authentication among many others.